# MESOSCALED PATTERNING IN NANOSTRUCTURED THIN FILMS/SILICATE GLASS SYSTEMS

K. Starbova[1], N. Starbov[1] and V. Tonchev[2*]

[1]Central Laboratory of Photoprocesses and [2]Institute of Physical Chemistry
Bulgarian Academy of Sciences, 1113 Sofia, Bulgaria

ABSTRACT

Various nanostructured thin solid films on silicate based glass plates commonly used as substrates in contemporary vacuum technologies are studied. Evolution of mesostructured spiral and spiral crack patterns as a result of shrinkage or crystallization processes is demonstrated. It is established that the surface hydrated layer of the glass is liable for the patterning observed. Evidence is presented supporting a mechanism of selforganisation phenomena in the systems studied. It is found that nano-sized sol gel silica thin film can be successfully used as a model system of the surface hydrated layer. Comparison of different thin solid films on silicate glass plates shows contour and dimensional similarity of memory patterns thus pointing out the occurrence of a general phenomenon.

INTRODUCTION

Silicate glass plates have been found in a broad spectrum of applications in the contemporary technologies as well as in fundamental research as substrates for deposition of compound, elemental and composite thin films by various physical and chemical techniques. The silicates possess a very long-term phase stability but mesoscaled ''memory'' effects are also observed in different thin film/silicate glass systems. That specific patterning has been easily recognized via optical imaging either in transmission or reflection mode. The incomplete crystallization at room temperature of nanostructured PVD antimony films on silicate glass below certain critical thickness can be regarded as an example of these phenomena [1, 2]. Relatively small number of active centres is found to catalyze the crystallization processes. The latter occurs in lateral direction resulting in circular patterns with strong increase of the optical reflectivity. The origin of the active centers and the film crystallization in the form of mesoscaled circular patterns is not discussed. Similar crystallization patterns that exhibit 2D spherulitic morphology have been observed in vapor grown p-nitrophenyl nitronyl nitroxide (p-NPNN) thin solid films on glass substrates [3]. As noted by the authors the low density of nucleation centers allows an accurate determination of the time evolution of the spherulitic radii. For small film thickness a non-linear dependence is found and it is explained considering the contribution of the interfacial energies at the beginning of the crystallization process [3].

Important for the development of cheep solar cells are the studies on crystallization of a-Si via Al induced layer exchange at relatively low temperatures in a-Si/Al/glass system. It is demonstrated that the amorphous to crystalline phase transition is initiated by a small number of active centers and proceeds laterally resulting in mesoscaled circular patterns, defined as large polysilicon grains. Optical transmission micrographs of time evolution of the polycrystalline Si demonstrate in all cases an incomplete crystallization but the role of glass substrate for the mesoscaled patterning was not subject of authors interest [4].

Thin layers obtained after drying of different inorganic precipitates display spiral cracks on a mesoscale. An unusual stress relaxation process is believed to be the reason for the observed symmetry-breaking fracturing mode. Computer simulations based on a coarse-grained model for fragmentation successfully reproduce the spirals. The latter are found to be logarithmic after fitting the experimental and simulation data. The authors predict that spiral cracks are not restricted to the studied compounds only [5]. Similar spiral patterns have been observed in multilayer coatings of nano-thick Mo and Si on silicon substrate as originally reported in [6]. The spirals are regarded as fracture patterns representing an example of though-thickness crack networks since generally thin films exhibit high levels of residual stress, up to several GPa, which are typically higher in compression compared to stretching.

---

[*] Corresponding author - e-mail:tonchev@ipc.bas.bg

Channel and surface patterns of mesoscale length comprising concentric circles, herringbones, fingerprints and hairpins have been observed in freestanding mesoporous silica films [7]. These patterns are considered as spatio-temporal silicified recordings of the polymerization and growth of a silicate liquid-crystal 2D seed emerging at the air/water interface. The existence of defects in the precursor confined by the surface is supposed to create a liquid-crystalline texture in the mesoporous silica film.

A study of E-glass fibers subjected to the action of oxalic, hydrochloric, nitric and sulfuric acids demonstrates generation of mesoscaled axial or spiral cracks on the fiber surface as depending on acid type and concentration [8]. The corrosion is primarily attributed to calcium and aluminum ion depletion. For that reason an ion-depletion-depth model is proposed to explain the mechanisms of fiber surface cracking. A smaller ion-depletion depth is accepted as a driving force for spiral crack formation as compared to that for axial cracks.

In all above mentioned studies of mesoscaled pattern formation the role of the glass substrate is not discussed. However, the crucial role of the surface hydrated gel layer is revealed [9-11] in nanoscaled a-C thin films on soda lime glass plates. In this case where mesoscopic spirals in the near surface glass region are decorated in a chemical way [9, 10]. Simultaneously, the presence of spiral cracks in the C-film itself after deliberate delamination of the latter from the glass substrate is demonstrated [10]. Besides, it is unambiguously shown that spiral morphogenesis and interactions follow some general trends of selforganization phenomena [9]. The role of alkaline oxides as guiding centers for spiral formation is demonstrated via experiments with model silicate glasses [11]. It seems that some universal features should construct the underlying mechanism of mesoscopic pattern formation.

In the present paper the results from experiments with various coatings on silicate glass plates used as substrates for film preparation are summarized. New examples of mesoscopic patterns are demonstrated expecting to contribute for understanding the mechanism of pattern formation.

**EXPERIMENTAL**

The present paper is based on investigations of different nanostructured thin films deposited via various coating techniques on Na-Ca-silicate glass.

The first object in this study represents Sb/Se bilayered thin film system prepared under high vacuum conditions on pre-cleaned glass plates by consecutive PVD deposition of Sb (43 nm) and Se (57 nm) in stoichiometric atomic proportion 2:3 between both elements. The samples were subjected to laser processing using CW $Ar^+$ laser operating at all lines (488 - 514 nm) in order to induce melting/solidification processes leading to synthesis of crystalline $Sb_2Se_3$ films. The irradiations were performed for 60 s at laser power of 0.8 W and 1.6 W. The early and advanced stages of laser synthesis of $Sb_2Se_3$ were imaged and recorded under optical microscope.

Specific morphological features found in nanostructured $Al_2O_3$ films that are partially delaminated from the glass surface regions will be shown. These samples were produced via high power e-beam evaporation from pure alumina tablets on stationary glass plates at a mean deposition rate of 0.3 nm/s. The heat flux released during this deposition process leads to substantial increase of the substrate temperature.

Applying quite different deposition technique sol-gel $SiO_2$ thin films were spin coated on same Na-Ca-silicate glass plates at room substrate temperature and using tetra ethyl ortho silicate (TEOS) as precursor. Low (10%) and high (40%) concentration of TEOS in ethanol were studied, keeping a constant hydrolysis degree (molar fraction $TEOS:H_2O=3$). After spin coating procedure single or bilayered gel films were obtained. In order to perform condensation process and to obtain solid films some of the gel smples were further subjected to non-isothermal drying for 20 min at $120^oC$. Optical and secondary electron imaging are applied for recording the morphologies obtained.

In order to simulate the heating processes in the experiments mentioned above pure Na-Ca-silicate glass substrates were irradiated by means of excimer $KrF^+$ pulsed laser ($\lambda = 248$ nm, $\tau = 40$ ns). This UV photon energy is absorbed within the substrates resulting in surface modification of the glass. The irradiations were performed with 100 pulses (repetition rate 10 Hz) at a laser energy density on the sample surface of 2.0 $J/cm^2$. The laser induced changes in the surface hydrated layer were decorated by means of short treatment in HF. After washing in distilled water and drying the samples were inspected in optical transmission mode the observed glass microstructure being recorded on photographs.

**RESULTS**

Fig.1 presents optical transmission micrographs of advanced stages (a, b) and of complete (c) laser induced synthesis of $Sb_2Se_3$. Spherulitic crystallization on small number active centers at low (b) and high (c) laser power is clearly seen. Moreover, specific circular or spiral-like image contrast is seen to superimpose on the spherulites in dark field mode (a) thus evidencing the presence of additional morphological features.

Fig.1

On Fig.2 a transmission optical micrograph of a delaminated from the substrate surface region of $Al_2O_3$ thin film is shown. Spiral-like cracks in the film are clearly seen their interaction strikingly reminding the selforganization phenomenon in a-C/silicate glass system mentioned above [9-11].

Fig.3 demonstrates HF etch patterns in the surface hydrated layer after excimer laser processing of Na-Ca-silicate glass substrate. The revealed morphological features are result of thermally induced processes in the glass subsurface region as a result of UV light absorption. An overall fractal-like crystallization is seen to coexist with spiral-like areas exhibiting a lower optical transmittivity. Most probably, the latter is due to Si-O bond breaking and separation of Si.

Fig.2, Fig.3 and Fig.4

Another set of experiments on $SiO_2$ gel films spin coated at high precursor concentration unambiguously show that the glass surface exhibits non-uniform wetting. The circular surface regions on Fig.4 are hydrophobic with respect to the solution used while in the remaining part a gel film is formed. Therefore, the coated gel silica films can be successfully used for decoration of both specific regions. The presence of guiding centers within the region free from sol-gel film is evident. The effect is less pronounced in $SiO_2$ gel films coated at low precursor concentration and becomes visible only after 4-fold consecutive coating of single gel films. As mentioned in the experimental part further drying of $SiO_2$ gel is necessary in order to obtain solid silica films. For these experiments two or three layers uniformly coated at low precursor concentrations on the glass substrate are subjected to non-isothermal treatment at $120^oC$. The results from the optical imaging revealed fascinating mesoscopic crack patterns accompanied by partial delamination from the glass surface as could be seen on Fig.5 (a, b). A saw-toothed spiral delamination (Fig.5a) is usually attributed to residual compressive stress relief and is an example of most common fracture patterning in thin films [12].

Fig.5 and Fig.6

Spaghetti-like fracture patterns together with strong delamination are observed in sol-gel $SiO_2$ films deposited by three-fold consecutive coating procedures (Fig.5b). In that cases a local decohesion between the individual single layers takes place (Fig.6) accompanied with a strong film shrinkage. The latter is more pronounced in lateral as compared to normal direction having in mind that the film thickness is of the order of 150 nm. Besides, initial stages of spiral formation reveal the presence of active centers on the glass surface.

In summary, the present study demonstrates a variety of mesoscale patterning in thin film/silicate glass systems obtained by different preparation and decoration techniques. Simultaneously, specific regions are visualized in the glass surface hydrated layer on a mesoscale.

**DISSCUSION**

It should be stressed that for all systems under investigation in the present study as well as for those published earlier [9-11] the silicate glass surface is the same. For that reason the following general conclusions could be made. The heat delivered to the glass surface by different routes could be considered to be the driving force of the observed patterning [9-11]. The heating/cooling rate is obviously of lesser importance having in mind the great difference between the strong non-equilibrium laser and the more or less quazi-equilibrium furnace heating. Spiral corrugations have been observed still in 19-th century in the outer surface of fulgurites obtained after a lightning hits sandy materials on the earth [13] as well as in Pyrex glass[14] and after shrinkage of silica gels [15]. Obviously, spiral formation could be attributed to the glass surface itself and more precisely to the surface hydrated gel layer rather than to the deposited thin films as usually considered until now. The changes in this surface layer should be strongly related to the parameters of the heat supplied during or after film coating. The resulting morphologies or patterning mode sharing a same mesoscopic length scale are specific for different thin films (spiral cracking, spherulite crystallization etc.) thus evidencing the great importance of their structure and properties for the observed phenomena. Additional conclusions are imposed with respect to silica gel films prepared by sol-gel technique on silicate glasses substrates. The contemporary technologies for preparation of high quality glass surfaces commonly comprise a sequence of wet procedures that could stabilize rather than eliminate this layer. In practice, the behavior of this tightly bound to the bulk glass silica gel film should be very similar to that of gel $SiO_2$ films. On this basis the latter could be considered as a model of the surface hydrated layer. The heat supplied to this layer should initiate effects of shrinkage and therefore the formation of crack patterns. Besides, crystallization process stimulated by the heat influx could also take place.

The presence of small number of active centers for mesoscopic patterning is attributed to the stabilizing oxides, as shown in our model experiments [11]. In the case of pure silica glasses these could be specific defect structures with higher water content. The absence of the mesoscopic patterning in the system a-C/silicate glass after deliberate sputtering of the surface layer under high vacuum conditions [11] could be regarded as the first evidence supporting the above conclusions.

ACKNOWLEDGEMENTS
The authors are highly obliged to B.Rangelov for the help during the preparation of the manuscript.

**Figure Captions**

**Figure 1**. Optical micrographs of CW Ar$^+$laser induced $Sb_2Se_3$ synthesis in 100 nm thick Sb/Se bilayered thin film on glass. Laser power 0.8 W (a, b) and 1.6 W (c); imaging performed in dark field mode (a), with polarized light (b), in transmission mode(c).

**Figure 2.** Optical transmission micrograph of delaminated from glass substrate surface 1000 nm thick $Al_2O_3$ thin film.

**Figure 3.** Optical transmission micrograph of excimer laser processed and HF etched surface of Na-Ca silicate glass.

**Figure 4.** Optical transmission micrograph of spin coated $SiO_2$ gel film on Na-Ca-silicate glass.

**Figure 5.** Optical micrographs in polarizing mode of bi- (a) and three- layered (b) sol-gel $SiO_2$ films on Na-Ca-silicate glass after 20 min non-isothermal treatment at 120$^o$C.

**Figure 6**. Scanning electron micrograph of three-layered sol-gel $SiO_2$ films after 20 min non-isothermal treatment at 120$^o$C.

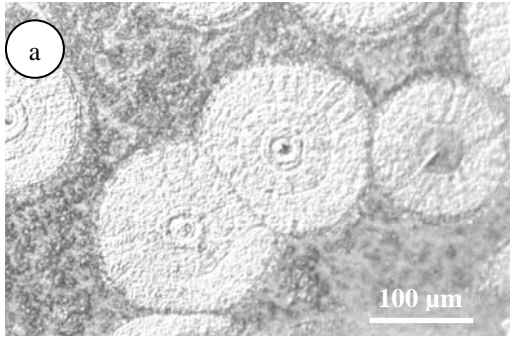
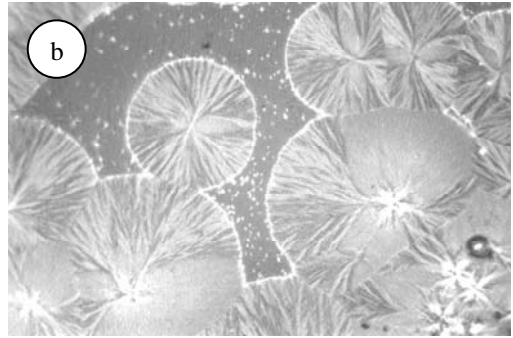
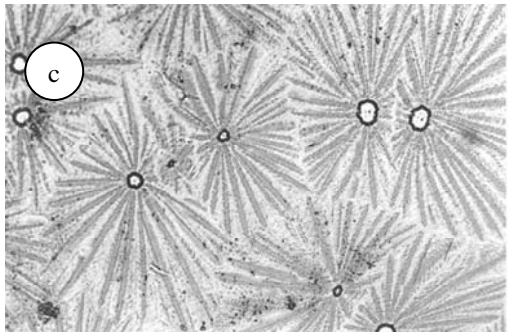

Fig.1 Optical micrographs of CW Ar*laser induced $Sb_2Se_3$ synthesis in 100 nm thick Sb/Se bilayered thin film on glass. Laser fluence 0.8 W (a,b) and 1.6 W (c); imaging performed in dark field mode (a), with polarized light (b), in transmission mode (c)

Mesoscaled Patterning in Nanostructured Thin Films/Silicate Glass Systems

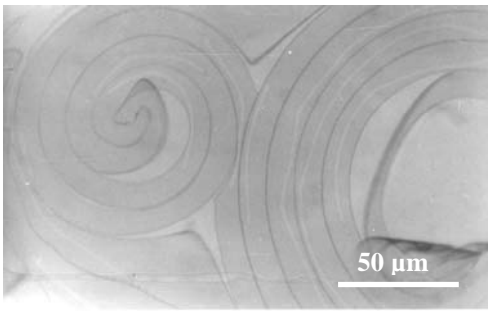

Fig.2. Optical transmission micrograph of delaminated from glass substrate surface 1000nm thick $Al_2O_3$ thin film

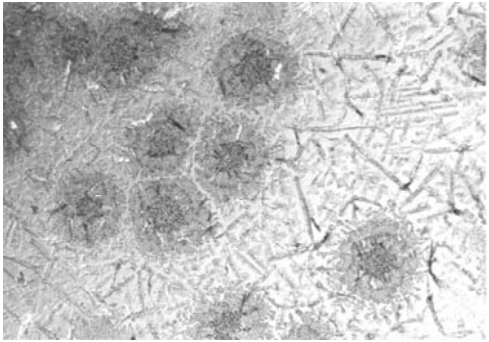

Fig.3 Optical transmission micrograph of excimer laser processed and HF etch surface of Na-Ca silicate glass

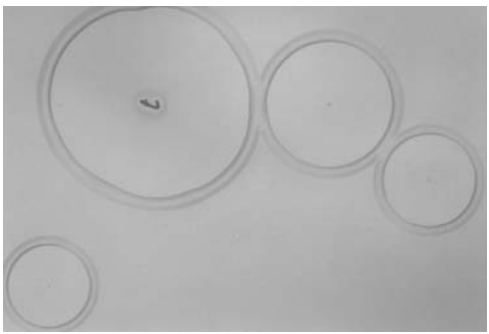

Fig.4. Optical transmission micrograph of spin coated $SiO_2$ gel film on Na-Ca-silicate glass

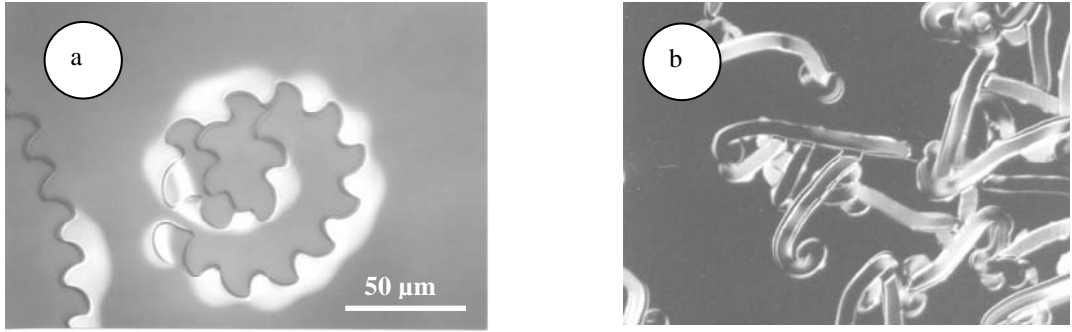

Fig.5. Optical micrographs in polarizing mode of bi- (a)and three layered (b)sol-gel $SiO_2$ films on Na-Ca-silicate glass after 20 min.non-isothermal treatment at $120^oC$.

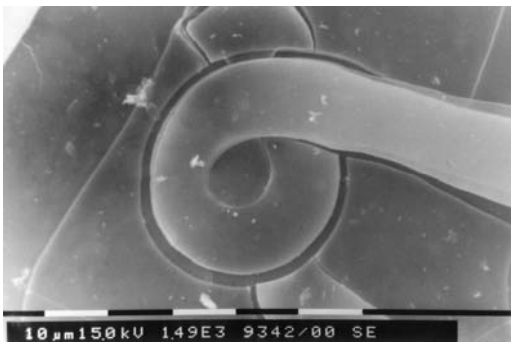

Fig.6 Scanning electron micrograph of three layered sol-gel $SiO_2$ films after 20 min.non-isothermal treatment at $120^oC$